\documentclass[aps,prb,twocolumn,superscriptaddress,amsmath,amssymb,showpacs]{revtex4}

\usepackage{graphicx}
\usepackage{dcolumn}
\usepackage{graphics}
\usepackage{amssymb}
\usepackage{bm}
\usepackage[ansinew]{inputenc}
\usepackage{placeins}
\usepackage[pdftex]{color}
\begin{document}

\title{Multiband electronic characterization of the complex intermetallic cage system Y$_{1-x}$Gd$_x$Co$_{2}$Zn$_{20}$}

\author{M. Cabrera-Baez}
\affiliation{CCNH, Universidade Federal do ABC (UFABC), Santo Andr\'e, SP, 09210-580 Brazil}

\author{A. Naranjo-Uribe}
\affiliation{Instituto de F\'{\i}sica, Universidad de Antioquia UdeA, Calle 70 No 52-21, Medell\'{\i}n, Colombia}

\author{J. M. Osorio-Guill\'en}
\affiliation{Instituto de F\'{\i}sica, Universidad de Antioquia UdeA, Calle 70 No 52-21, Medell\'{\i}n, Colombia}

\author{C. Rettori}
\affiliation{CCNH, Universidade Federal do ABC (UFABC), Santo Andr\'e, SP, 09210-580 Brazil}
\affiliation{Instituto de F\'{\i}sica ``Gleb Wataghin", UNICAMP, Campinas, SP, 13083-859 Brazil}

\author{M. A. Avila}
\affiliation{CCNH, Universidade Federal do ABC (UFABC), Santo Andr\'e, SP, 09210-580 Brazil}
\date{\today}

\begin{abstract}
A detailed microscopic and quantitative description of the electronic and magnetic properties of Gd$^{3+}$-doped YCo$_{2}$Zn$_{20}$
single crystals (Y$_{1-x}$Gd$_{x}$Co$_{2}$Zn$_{20}$: (0.002 $\lesssim x \leq $ 1.00) is reported through a combination of temperature-dependent electron spin resonance (ESR),
heat capacity and $dc$ magnetic susceptibility experiments, plus first-principles density functional theory (DFT) calculations.
The ESR results indicate that this system features an \emph{exchange bottleneck} scenario wherein various channels for the spin-lattice relaxation 
mechanism of the Gd$^{3+}$ ions can be identified via exchange interactions with different types of conduction electrons at the Fermi level.
Quantitative support from the other techniques allow to extract the exchange interaction parameters between the localized magnetic moments of the
Gd$^{3+}$ ions and the different types of conduction electrons present at the Fermi level ($J_{fs}$, $J_{fp}$ and $J_{fd}$).
Despite the complexity of the crystal structure, our combination of experimental and electronic structure data establish GdCo$_{2}$Zn$_{20}$ as a
model RKKY system by predicting a Curie-Weiss temperature $\theta_{C} = -1.2(2)$~K
directly from microscopic parameters, in very good agreement with the bulk value from magnetization data.
The successful microscopic understanding of the electronic structure and behavior for the two end compounds YCo$_{2}$Zn$_{20}$ and
GdCo$_{2}$Zn$_{20}$ means they can be used as references to help describe the more complex electronic properties of related materials.
\end{abstract}


\maketitle

\section{Introduction}

The understanding of the physical properties of complex materials with cage-like structures, such as the family RT$_{2}$Zn$_{20}$
(R = rare earth, T = transition metal), have attracted the attention of many researchers focused in condensed matter physics. 
Among other aspects, these intermetallic cage compounds generate interest due to the different types of electronic and magnetic behaviors governed by ``naturally diluted'' rare earth ions.\cite{Jia0,Jia}
The magnetic versatility associated with the 4$f$ electrons goes from weakly correlated Pauli-like paramagnetic behavior (Lu$^{3+}$), to hybridization with conduction electrons (Yb$^{3+}$) and peculiar interactions of local magnetic moments (Gd$^{3+}$) with conduction electrons, as a few examples.

\begin{figure}[!htb]
\begin{center}
\includegraphics[width=92mm,keepaspectratio]{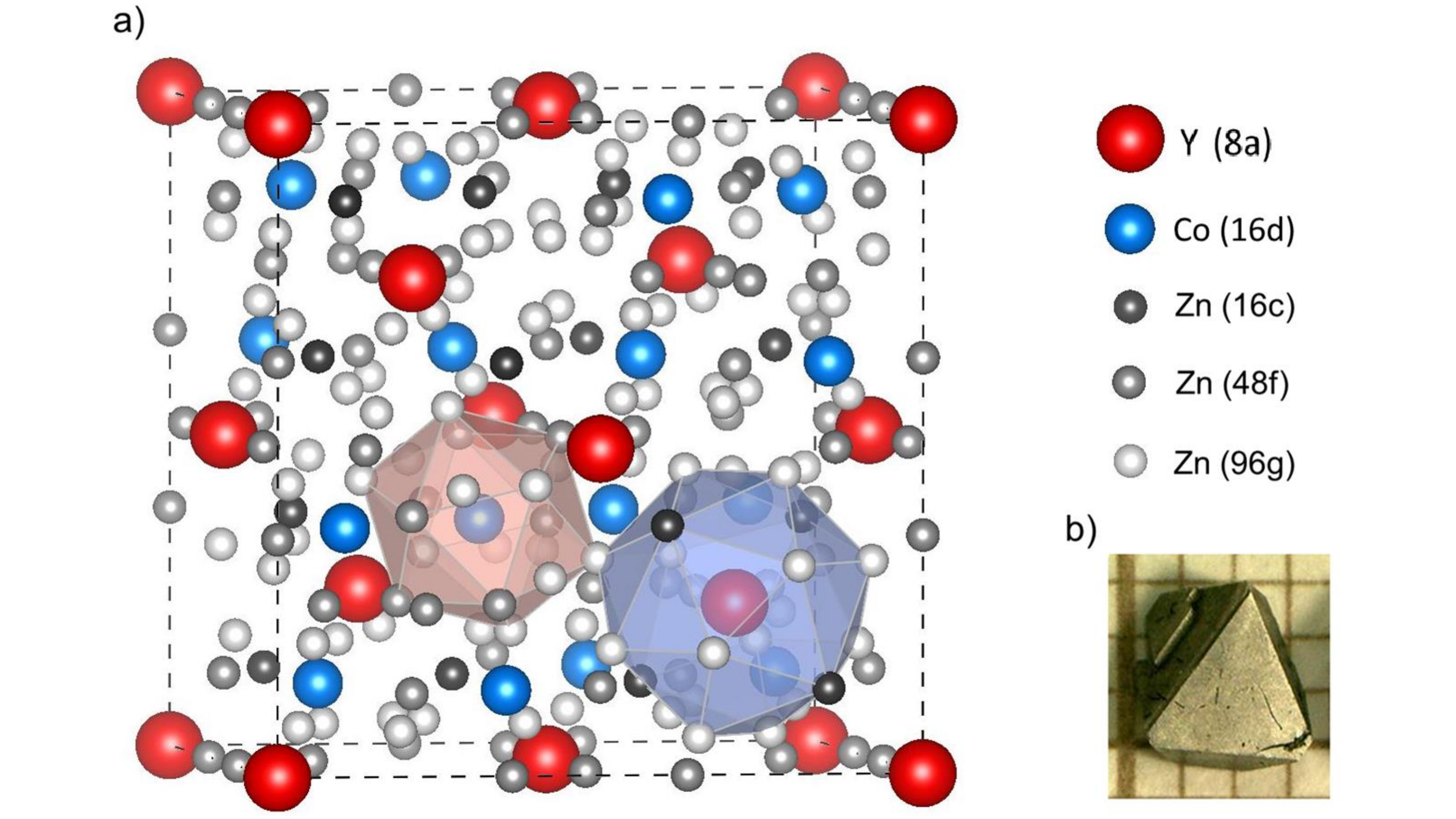}
\end{center}
\vspace{-0.6cm} \caption{(a) Conventional unit cell of YCo$_{2}$Zn$_{20}$ (space group $Fd\bar{3}m$).
Y, Co and Zn ions are represented by red, blue and gray balls, respectively.
The CN-16 Frank-Kasper polyhedron around the Y ions and the polyhedron around Co ions are highlighted in pink and violet, respectively.
(b) Typical crystal of YCo$_{2}$Zn$_{20}$ showing a [111] facet.}\label{structure}
\end{figure}

Despite the apparent complexity (184 atoms per conventional unit cell arranged in different types of cages), their crystallographic structure can be broken down into surprisingly simple sub-units, which allows clean analyzes and interpretations of experimental results, and sets this family as an excellent model system for several physical problems. 
They adopt a cubic CeCr$_{2}$Al$_{20}$-type structure (space group: $Fd\bar{3}m$),\cite{Nasch} in which the R and T ions occupy their own unique crystallographic sites ($8a$ and $16d$, respectively).
The Zn ions form the cage structure by occupying three inequivalent crystallographic sites ($96g$, $48f$ and $16c$).
If we consider the sub-structure in terms of the nearest neighbors and the next nearest neighbors, the R and T ions are fully surrounded by shells formed by Zn ions, leaving a shortest R-R spacing of $r \approx 6$~\AA.
The R ions are thus isolated in Frank-Kasper cages formed by 16 Zn ions as exemplified in Fig.~\ref{structure}.

The observation of radically different magnetic behaviors such as the contrast between low-temperature antiferromagnetic order ($T_N\sim5.7$~K)\cite{Jia2} in GdCo$_{2}$Zn$_{20}$  and the high temperature ferromagnetic order ($T_C\sim86$~K)\cite{Jia2} in GdFe$_{2}$Zn$_{20}$ hints at peculiar magnetic couplings governed by the Ruderman-Kittel-Kasuya-Yosida (RKKY) interaction.
In the case of Y-based compounds, largely different types of behavior are also found.
YFe$_{2}$Zn$_{20}$ has been described as a ``nearly ferromagnetic Fermi sea" compound\cite{Jia3} because it is near the Stoner limit, in contrast to YCo$_{2}$Zn$_{20}$ with more conventional metallic behavior.
Moreover, within this family all the Yb-based compounds described so far have presented heavy fermion behavior,\cite{Torikachvili} as evidenced by the enhanced Sommerfeld coefficients, reaching $\gamma \approx 7900$~mJ/mol.K$^2$ for YbCo$_{2}$Zn$_{20}$.
All of these observations point to the need to investigate the electronic structure,\cite{Tanaka} particularly around the Fermi level, and describe the electronic interactions in detail to better understand the electronic and magnetic behaviors.

In order to conduct this task in a tractable manner, we have chosen an initial focus on the weakly correlated compound YCo$_{2}$Zn$_{20}$ that features Pauli like paramagnetism, metallic transport and a Sommerfeld coefficient of $\gamma = 18.3$~mJ/mol.K$^2$,\cite{Jia,Jia2} as an appropriate host for a microscopic study using Gd$^{3+}$ ions as an electron spin resonance (ESR) probe.
ESR of rare earth ions diluted in metallic hosts is a useful local technique to investigate microscopic properties of materials, since it directly probes the localized magnetic moments and the nature of the interactions with their neighbors.\cite{Taylor, Barnes}
The metallic and nonmagnetic YCo$_{2}$Zn$_{20}$ host doped with Gd is an excellent model system to study the Gd$^{3+}$ spin-lattice relaxation, associated with the conduction electrons (\emph{ce}) spin-flip scattering mechanism due to the exchange interaction between the localized magnetic moment and the \textit{ce}.
The Hasegawa-Korringa model\cite{Hasegawa, Korringa} for the spin-lattice relaxation has been carefully discussed and applied in previous studies of ESR for Gd$^{3+}$ in the intermetallic compounds like LaAl$_2$ \cite{Davidov1}, LuAl$_2$ \cite{Rettori1} and in elemental Al.\cite{Rettori2}
We have recently applied the same technique to investigate the electronic structure of the superconductor YIn$_3$.\cite{Michael}

In this work we show that the ESR spectra of Gd$^{3+}$ in YCo$_{2}$Zn$_{20}$ ($0.001\lesssim x \leq 1.00$) presents a Gd$^{3+}$ concentration-dependent thermal broadening of the linewidth and $g$-shift.
This reveals the existence of the \textit{exchange bottleneck} effect in this compound, that can be tuned by the concentration of Gd$^{3+}$.
By combining the ESR results with heat capacity, magnetic susceptibility and band structure calculations, we extract the exchange parameters of the 
interaction between Gd$^{3+}$ and the $s$, $p$, and $d$ $ce$ present at the Fermi level of YCo$_{2}$Zn$_{20}$.
We are then able to establish a clear correlation of these microscopic parameters with the RKKY interaction.
This in turn offers a better understanding of the peculiar ``Fermi sea'' present in the system, which has lead to the magnetic anomalies found in this family.

\section{Experimental and computational details}

\begin{figure}[!ht]
\begin{center}
\includegraphics[width=85mm,keepaspectratio]{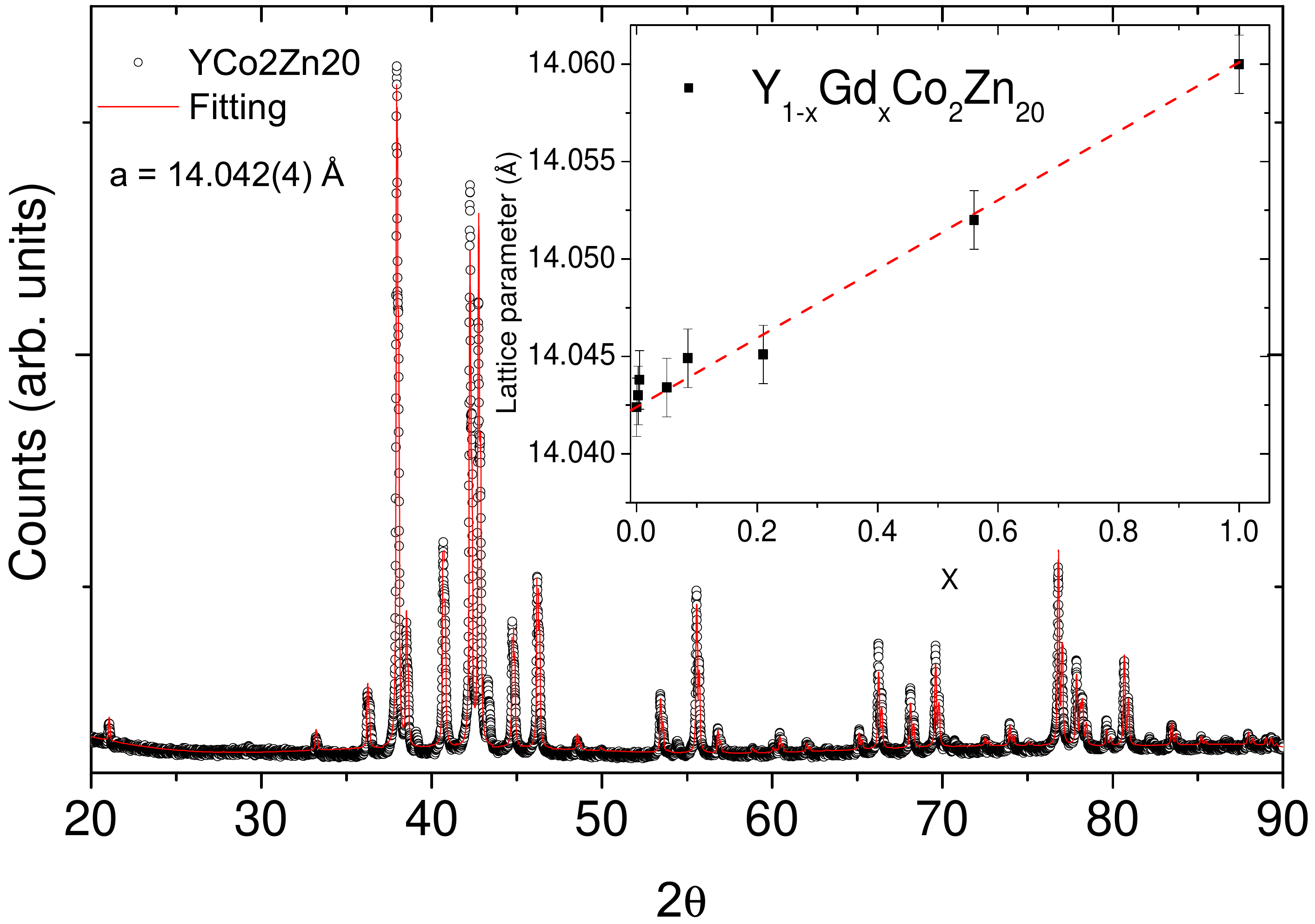}
\end{center}
\caption{XRD pattern of YCo$_{2}$Zn$_{20}$. Inset, Lattice parameter for the Y$_{1-x}$Gd$_{x}$Co$_{2}$Zn$_{20}$ (0.001 $\lesssim x \leq $ 1.00) 
compounds.}.\label{XRD}
\end{figure}

Batches of Y$_{1-x}$Gd$_{x}$Co$_{2}$Zn$_{20}$ (0.001 $\lesssim x \leq $ 1.00) single crystals were grown by the self-flux method\cite{Canfield,Raquel} using excess Zn.
The starting reagents were 99.9$\%$ Y, 99.9$\%$ Co, 99.9$\%$ Gd and 99.9999$\%$ Zn (Alfa-Aesar).
Initial ratios of elements were 1:2:47 for the pure ternaries Y:Co:Zn and Gd:Co:Zn, or 1-x:x:2:47 for the pseudo-quaternaries Y:Gd:Co:Zn, based on previously reported growths of the ternary compounds.\cite{Jia2}
The elemental mixtures were sealed in an evacuated quartz ampoule and placed in a box furnace for the temperature ramping.
Crystals were grown by slowly cooling the melt between 1100 $^\circ$C and 600 $^\circ$C over 100 h.
At 600 $^\circ$C the ampoules were removed from the furnace, inverted and placed in a centrifuge to spin off the excess flux.
The separated crystals are typically polyhedral, $\sim3$~mm or larger and manifest clear, triangular [111] facets (Fig.~\ref{structure}). 
The Gd concentrations were estimated based on the effective moments per formula unit extracted from fits of magnetic susceptibility measurements.
Powder x-ray diffraction on crushed crystals was used to ascertain the CeCr$_{2}$Al$_{20}$-type structure\cite{Nasch} as exemplified for YCo$_{2}$Zn$_{20}$ in Fig.~\ref{XRD}.
The refined lattice parameter of $a = 14.042(4)$~\AA~is in good agreement with the literature.\cite{Jia0}
The inset shows that the refined lattice parameter increases linearly with Gd concentration, as expected by Vegard's law.\cite{Vegard} 

For the ESR experiments, single crystals were crushed into fine powders of particle size greater than 100 $\mu$m, corresponding to average grain size (\emph{d}) larger than the skin depth ($\delta$), $\lambda = d/\delta \gtrsim 10$.
We have noticed that experiments on as-grown single crystals should be carried out with caution, because strong resonances similar to those reported by Ivanshin \emph{et al.}\cite{Ivanshin} are frequently observable, and likely due to residual Co surface contamination since that particular signal disappears after removing the as-grown crystal surfaces.
The X-Band ($\nu \approx9.4$~GHz) ESR experiments were carried out in a conventional CW Bruker-ELEXSYS 500 ESR spectrometer using a TE$_{102}$ cavity.
The sample temperature was changed using a helium gas-flux coupled to an Oxford temperature controller.
The specific heat ($C_p$) and magnetic susceptibility ($\chi=M/H$) measurements were performed on Quantum Design PPMS and SQUID-VSM platforms, respectively, using their standard procedures.

The ground state crystal structures were calculated using spin-polarized first-principles density functional theory (DFT), using the PBEsol exchange-correlation functional.\cite{Perdew}
The Kohn-Sham equations were solved using the projector augmented plane-wave (PAW) method as implemented in the VASP code.\cite{Kresse1, Kresse2}
The PAW atomic reference configurations are: 4$s^{2}$4$p^{6}$4$d^{1}$5$s^{2}$ for Y, 5$s^{2}$5$p^{6}$4$f^{7}$5$d^{1}$6$s^{2}$ for Gd, 3$s^{2}$3$p^{6}$3$d^{7}$4$s^{2}$ for Co and 3$p^{6}$3$d^{10}$4$s^{2}$ for Zn,
where only electrons treated as valence electrons are explicitly enumerated.
The energy cut-off in the plane-waves expansion is 507.5 eV, where the total energy has been converged to1 meV/unit cell. All structural parameters, lattice constants and atomic positions for each calculated compound have been optimized by simultaneously minimizing all atomic forces and stress tensor components via a conjugate gradient method.
Successive full-cell optimizations adapting basis vectors have been conducted until the unit cell energies and structural parameters were fully converged.
Brillouin-zone integration has been performed on a Monkhorst-Pack $12\times12\times12$ $\mathbf{k}$-point grid with a Gaussian broadening of 0.01 eV for full relaxation (ionic forces are converged to 0.1 meV/\AA).
Then, we used the relaxed crystal structures to calculate the total and partial density of states (DOS), the dispersion relations and the Fermi surfaces using the full-potential augmented-plane wave method with local orbitals.~\cite{Dewhurst}
The muffin-tin (MT) radii of Y, Gd, Co and Zn are set to $R_{\mathrm{MT}} = 2.6, 2.8138, 2.2804$ and 2.2804 a.u., respectively.
The parameter $R_{\mathrm{MT}} |\mathbf{G}+\mathbf{k}|_{\mathrm{max}}$ governing the number of plane waves has been converged to 9.0.
The irreducible wedge of Brillouin zone is sampled with a $20\times20\times20$ uniformly spaced $\mathbf{k}$-point grid.
Fermi surfaces were plotted using the XCrysDen package.\cite{Kokalj}

\section{Experimental Results}

\begin{figure}[!ht]
\begin{center}
\includegraphics[width=85mm,keepaspectratio]{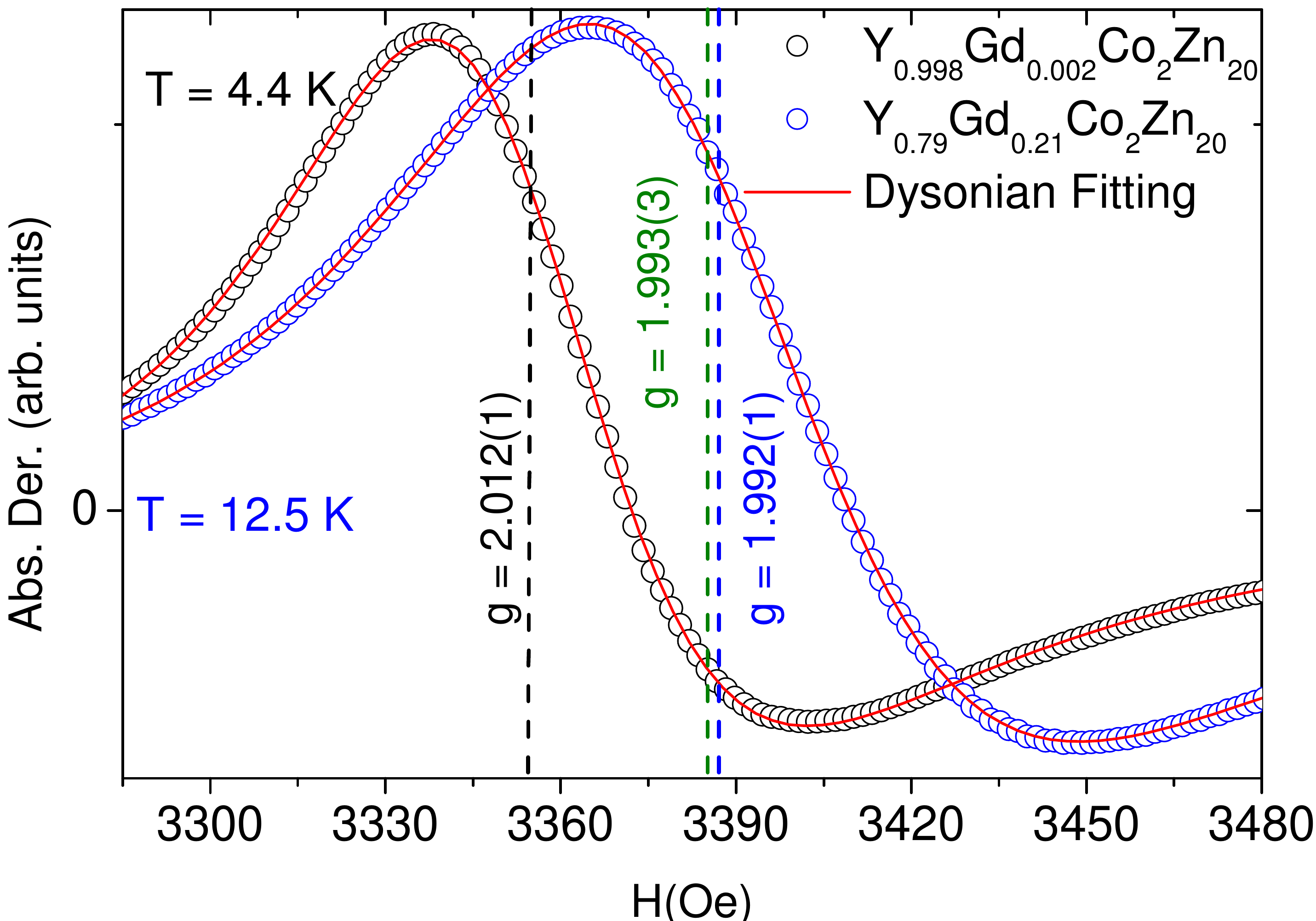}
\end{center}
\vspace{-0.7cm} \caption{Gd$^{3+}$ ESR spectra of Y$_{1-x}$Gd$_{x}$Co$_{2}$Zn$_{20}$ for $x\approx0.002$ at $T$ = 4.4 K and $x\approx0.21$ at $T$ = 12.5 K for a microwave power of $P_{\mu\omega}\approx2$~mW.}\label{dysonian}
\end{figure}

Figure~\ref{dysonian} displays the Gd$^{3+}$ ESR spectra in Y$_{1-x}$Gd$_{x}$Co$_{2}$Zn$_{20}$ for $x \approx 0.002$ and $x \approx 0.21$ at 4.4~K and 12.5~K, respectively, and microwave power of $P_{\mu\omega} \approx2$~mW.
These ESR spectra show different resonance magnetic fields than that of Gd$^{3+}$ in insulators, for which a resonance field of $H_0 = 3386(4)$~Oe and $g$-value of 1.993(2) are well established.\cite{Abragam}
It is evident that for the low concentration sample ($x = 0.002$) the resonance is shifted toward a lower field (higher $g$-values) compared to that of the higher concentration sample ($x = 0.21$).

The observed ESR spectra of Gd$^{3+}$ localized magnetic moments in Y$_{1-x}$Gd$_{x}$Co$_{2}$Zn$_{20}$ will be analyzed according to the generally accepted approach where, at resonance, the microwave absorption in a metal is given by the Dyson theory in the diffusionless limit, $A/B\approx 2.6$.\cite{Feher,Dyson}
In this limit, for particles larger than the skin depth, the ESR spectra reduce to a simple admixture of absorption $(\chi'')$ and dispersion $(\chi')$ of Lorentzian lineshapes.\cite{Feher,Dyson}
The derivative of this admixture is given by:

\begin{eqnarray}
\frac{d\left[(1-\alpha)\chi''+\alpha\chi'\right]}{dH}&=&\chi_{0}H_{0}\gamma_e^{2}
T_{2}^{2}\Biggr[\frac{2\left(1-\alpha\right)x}{\left(1+x^{2}\right)^{2}}\nonumber\\&+&\frac{\alpha\left(1-x^{2}\right)}{\left(1+x^{2}\right)^{2}}\Biggl]\\\nonumber
\\ x&=&\left(H_{0}-H\right)\gamma_e T_{2},\nonumber
\label{eq2}
\end{eqnarray}

where $H_{0}$ and $H$ are the resonance and the applied fields respectively, $\gamma_e$ is the electron gyromagnetic ratio, $T_2$ the spin-spin relaxation time, $\alpha$ the admixture of absorption $(\alpha=0)$ and dispersion $(\alpha=1)$ and $\chi_{0}$ the paramagnetic contribution from the static susceptibility.
It is usually accepted\cite{Dyson} that for diluted magnetic moments in a metallic host $T_1 \approx T_{2}$, where $T_1$ is the spin-lattice relaxation time.\cite{Abragam,Poole}
Therefore, the fitting of the experimentally observed ESR absorption lines to Eq.~1 allows the extraction of the two most relevant ESR parameters, i.e., the $g$-value from the resonance condition, $h\nu = g\mu_B H_{0}$, and the linewidth $\Delta H = 1/\gamma_e T_2$.

\begin{figure}[!ht]
\begin{center}
\includegraphics[width=85mm,keepaspectratio]{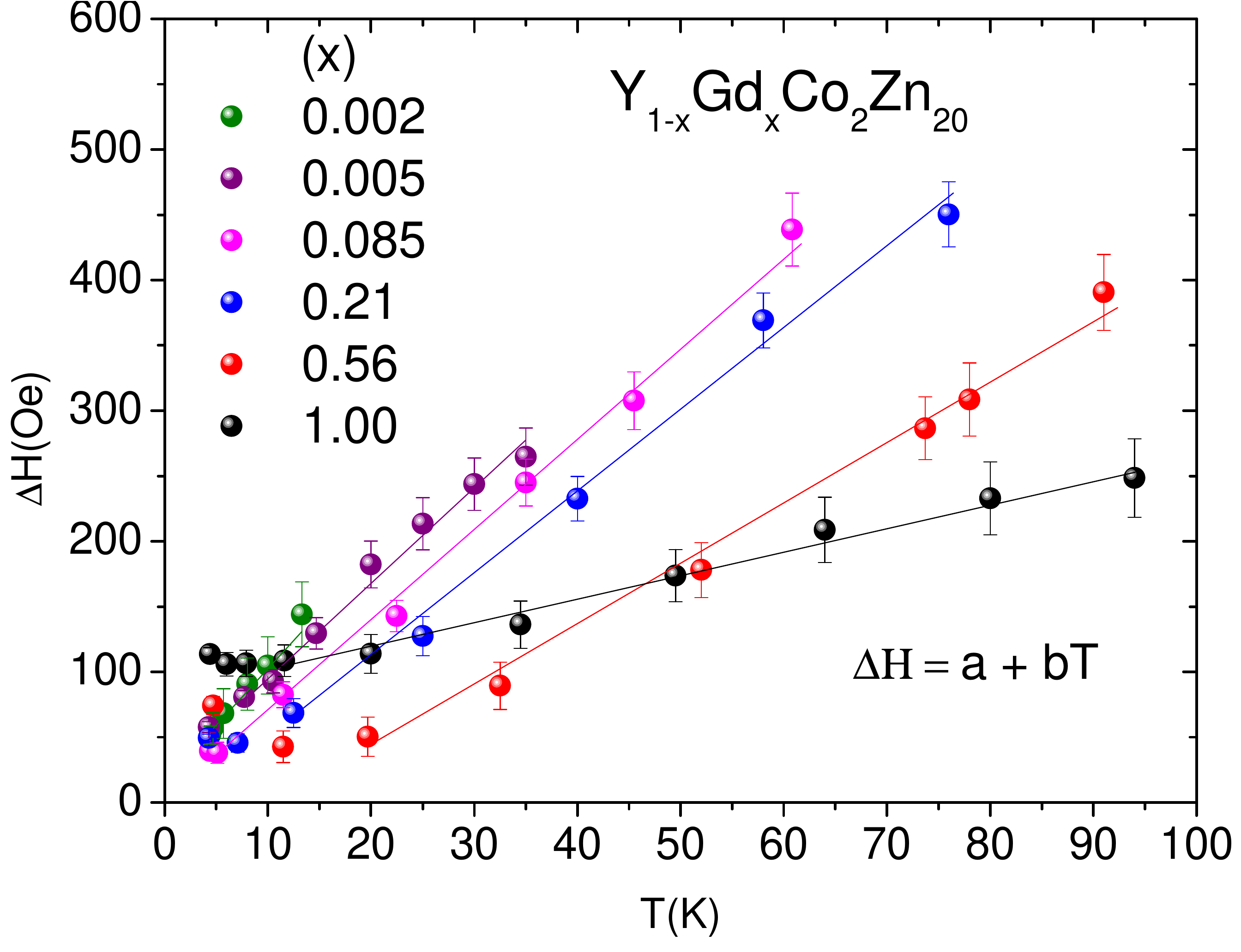}
\end{center}
\caption{$T$-dependence of the Gd$^{3+}$ ESR linewidth, $\Delta H$, in Y$_{1-x}$Gd$_{x}$Co$_{2}$Zn$_{20}$ for $0.001 \lesssim x \leq 1.00$.}\label{linewidth}
\end{figure}

\begin{figure}[!ht]
\begin{center}
\includegraphics[width=85mm,keepaspectratio]{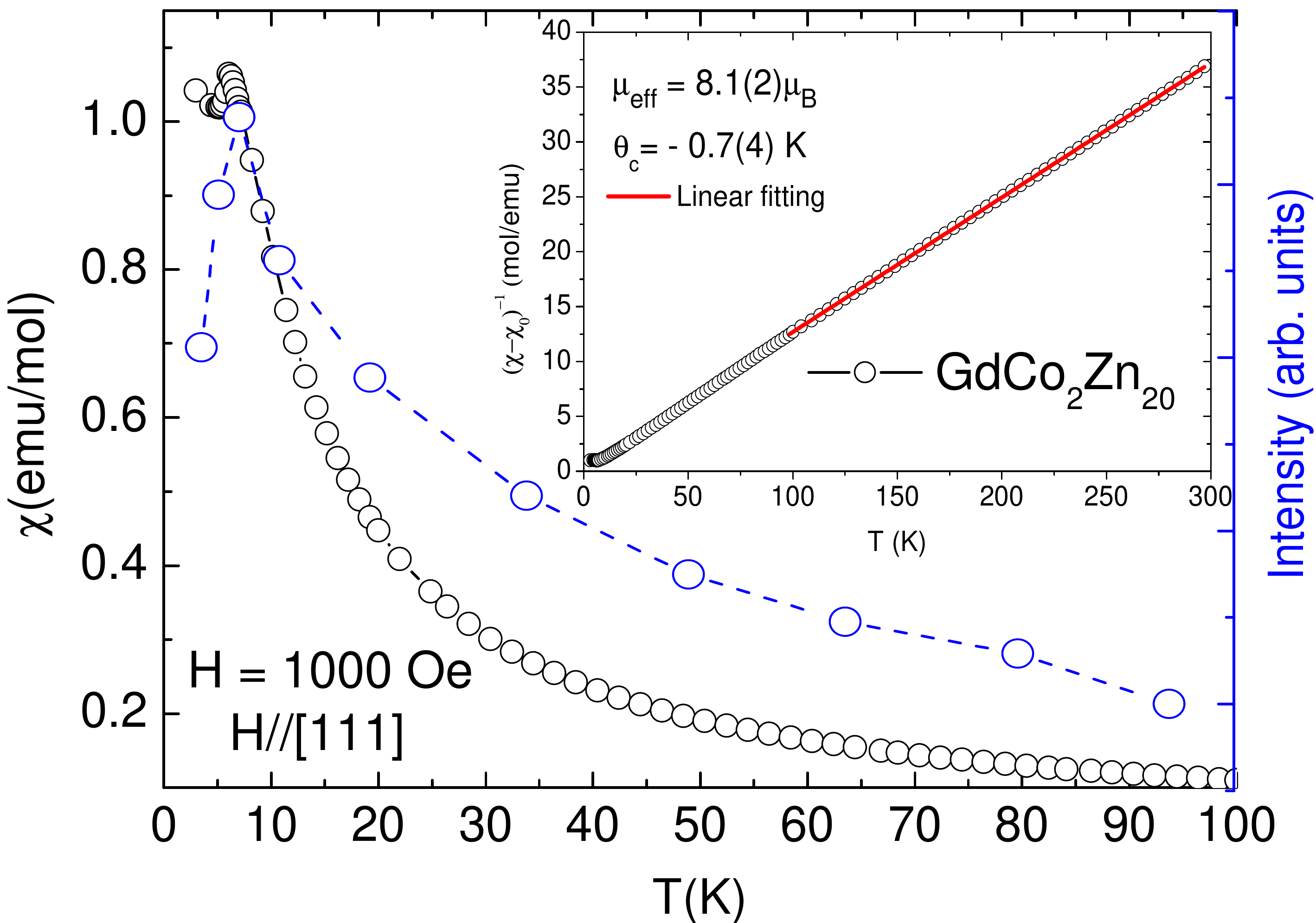}
\end{center}
\caption{$T$-dependence of the $dc$ magnetic susceptibility and the Gd$^{3+}$ ESR intensity for GdCo$_{2}$Zn$_{20}$.
Both experiments show the magnetic paramagnetic-antiferromagnetic transition at $T_N \approx5.7$~K.
The inset shows the inverse magnetic susceptibility and linear fit results.}\label{suscept}
\end{figure}

Figure~\ref{linewidth} displays the $T$-dependence of the Gd$^{3+}$ ESR linewidth, $\Delta H$, in Y$_{1-x}$Gd$_{x}$Co$_{2}$Zn$_{20}$ for $0.002 \lesssim x \leq 1.00$ at a microwave power of $P_{\mu\omega} \approx2$~mW.
The broadening of $\Delta H$ at low temperature for the high concentration samples is presumably originated by the interaction between randomly distributed Gd$^{3+}$ magnetic moments, that cause an inhomogeneous local field.
However, for the stoichiometric GdCo$_{2}$Zn$_{20}$ this disorder should be absent and the low temperature broadening of $\Delta H$ less pronounced (black symbols in Fig.~\ref{linewidth}).
This allows the drop in the magnetic susceptibility at $T_N$ to become evident in the ESR intensity (see Fig.~\ref{suscept}).
The high $T$-dependence of $\Delta H$ follows the linear behavior $\Delta H = a + bT$ where the $a$ parameter represents the residual linewidth, $\Delta H_{0}$, and $b=d(\Delta H)/dT$ the Korringa-like relaxation rate.
The extracted $a$ and $b$ values are given in Table I, together with the obtained $g$-values at $T\approx$ 10 K for the studied samples.

\begin{figure}[!ht]
\begin{center}
\includegraphics[width=85mm,keepaspectratio]{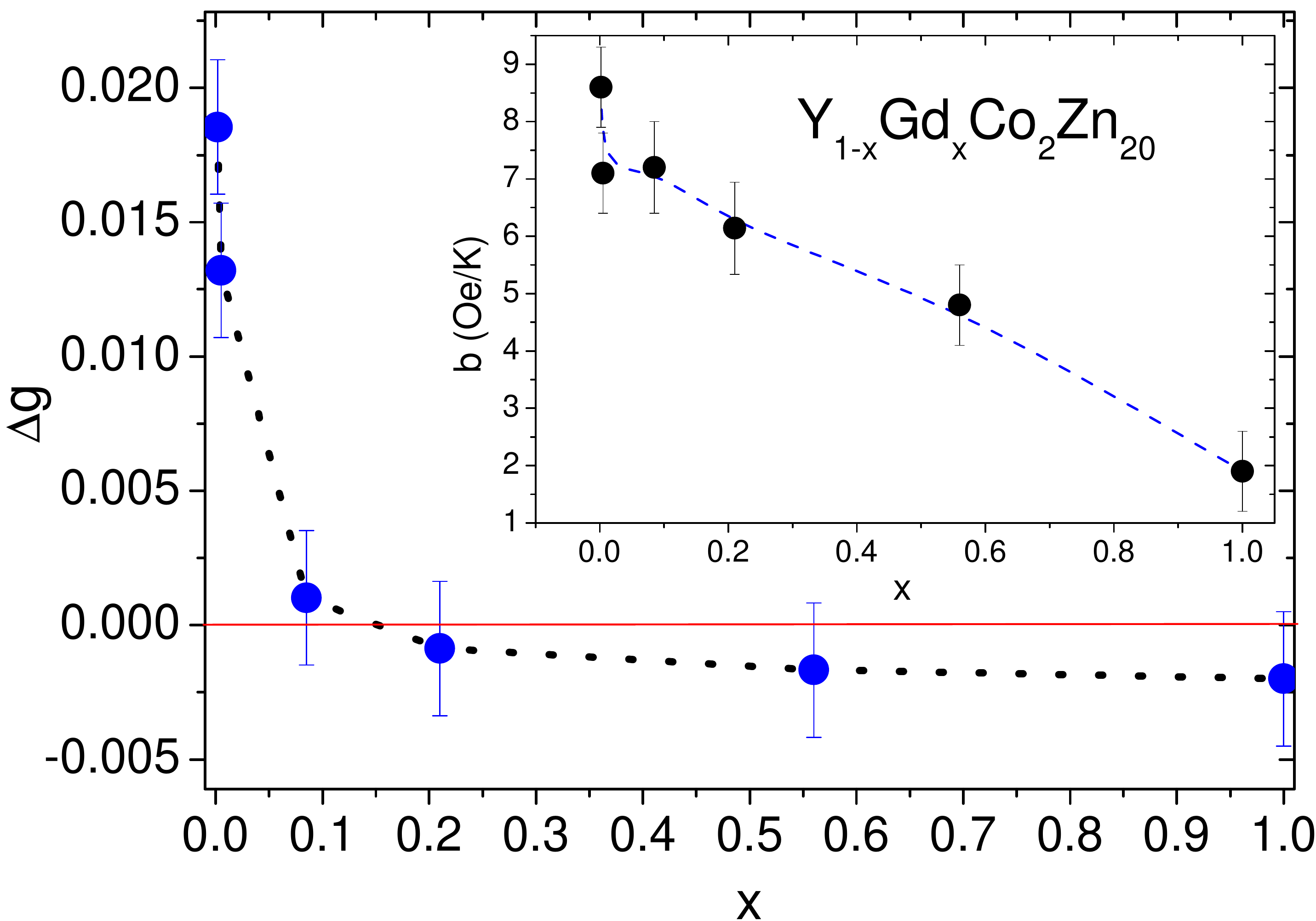}
\end{center}
\caption{Gd concentration dependence of the $g$-shift ($\Delta g$ = $g$ - 1.993(2)) and in the Inset the thermal broadening of the linewidth, $b$, for the Y$_{1-x}$Gd$_x$Co$_2$Zn$_{20}$ system.
The dashed lines are guides for the eye.}\label{deltag}
\end{figure}

Figure~\ref{deltag} displays the Gd concentration dependence of the $g$-shift ($\Delta g$ = $g$ - 1.993(2)) and in the inset the thermal broadening of the linewidth, $b$.
The general trends of the data presented in Fig.~\ref{deltag} is characteristic of an \textit{exchange bottleneck} phenomenon, where the $ce$ relaxation to the Gd$^{3+}$ localized magnetic moment (Overhauser relaxation) overcomes the $ce$ spin-lattice relaxation.

\begin{figure}[!h]
\begin{center}
\includegraphics[width=85mm,keepaspectratio]{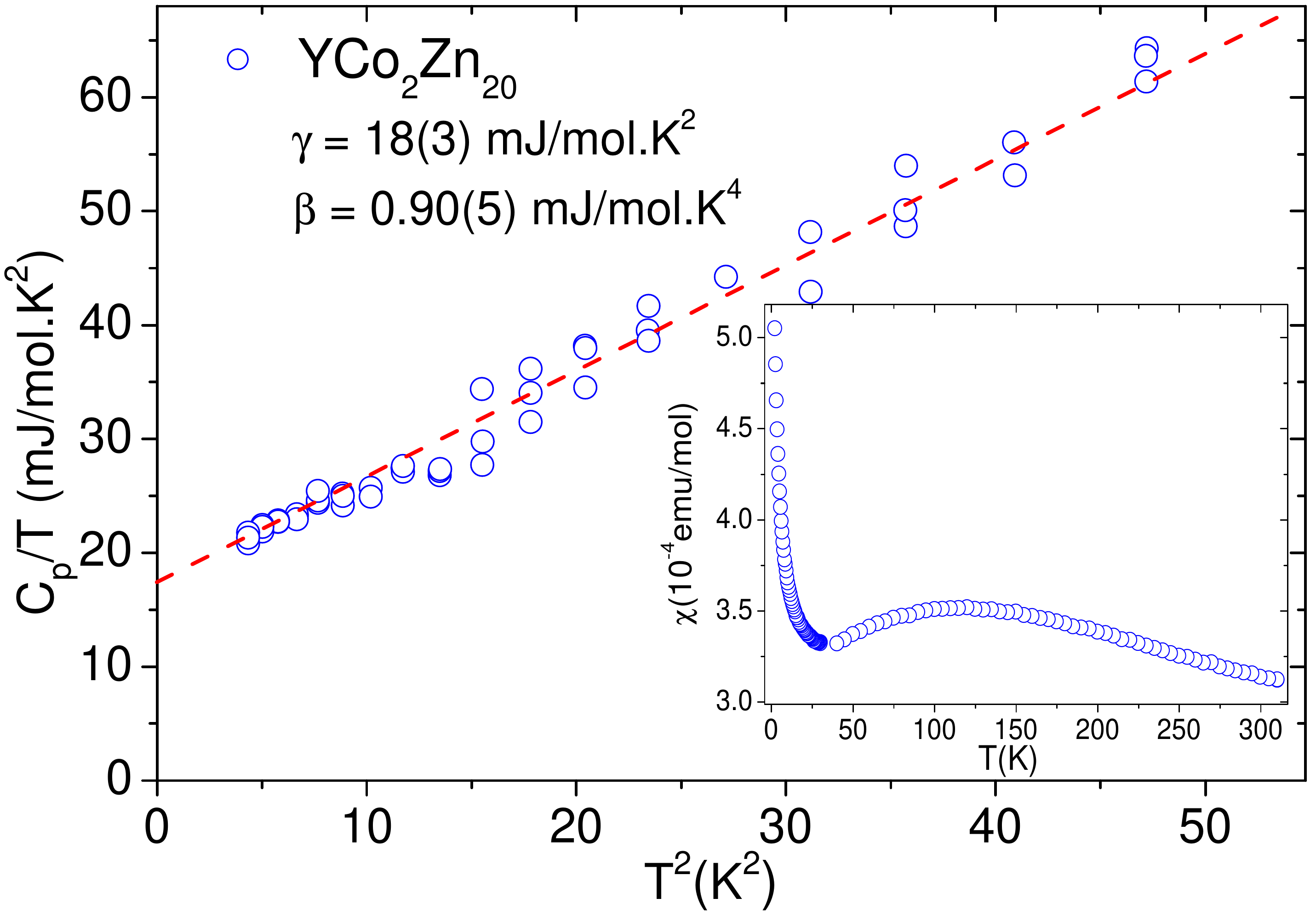}
\end{center}
\vspace{-0.7cm} \caption{Low temperature specific heat, $C_p/T$, showing a linear behavior with a Sommerfeld coefficient of $\gamma = 18(3)$~mJ/mol~K$^2$ and a Debye temperature of $\Theta_{D}$ = 370(7) K for YCo$_{2}$Zn$_{20}$.
The inset zooms in on the weak $T$-dependence of the magnetic susceptibility, $\chi(T)$, for YCo$_{2}$Zn$_{20}$}\label{heat}
\end{figure}

We now focus on bulk thermodynamic measurements which, together with the band structure calculations presented in the following section, provide support for a proper quantitative analysis of the ESR results.
Figure~\ref{heat} shows the low temperature linear behavior of $C_p/T$ as a function of $T^{2}$ leading to a Sommerfeld coefficient of $\gamma = 18(3)$~mJ/mol.K$^2$ and a Debye temperature of $\Theta_{D}$ = 370(7) K for the YCo$_{2}$Zn$_{20}$ compound. 
Our obtained value of $\gamma$ is in agreement with the previously reported value of 18.3~mJ/mol.K$^2$.\cite{Jia2} 
The inset of Fig.~\ref{heat} zooms in on the weak $T$-dependence of the magnetic susceptibility, $\chi(T)$, for YCo$_{2}$Zn$_{20}$.
The dome-like feature at $T \approx$ 120~K may be due to $T$-dependent $d$-$ce$ at the Fermi surface (see Fig.~\ref{DOS} below) and the small upward tail at low temperature (which has no influence in the analyzes that follow) due to residual rare-earth magnetic impurities from the 99.9\% Y reagents used in our samples.

Fig.~\ref{suscept} presents the $T$-dependence of the magnetic susceptibility of GdCo$_{2}$Zn$_{20}$ where the antiferromagnetic order is seen at a Neel temperature of $T_N \approx 5.7$~K, with an effective magnetic moment $\mu_{eff} = 8.1(2)$~$\mu_B$, comparable to that of Gd$^{3+}$ ions ($\mu_{eff}=7.94$~$\mu_B$) and a paramagnetic Curie temperature $\theta_{C} = -0.7(4)$~K.
Note that the paramagnetic Curie temperature $\theta_{C}$ is the same as $T_{C}$ in the Weiss molecular field theory.
Fig.~\ref{suscept} also shows that the ESR signal integrated intensity follows the trends of the magnetic susceptibility, crossing the paramagnetic-antiferromagnetic transition at approximately the same Neel temperature.
This is in itself a rare observation because usually, around the ordering temperature of a magnetic transition, the ESR signal is lost due to a strong broadening of the resonance. 

The magnetic susceptibility for all our Y$_{1-x}$Gd$_{x}$Co$_{2}$Zn$_{20}$ samples were fitted to a Curie-Weiss law using the effective magnetic moment of $\mu_{eff}=7.94\,\mu_B$ for the Gd$^{3+}$ ions.
From these fittings the Gd concentrations were estimated and their values are listed in Table~\ref{tab1}.

\section{Analysis and Discussion}

The low-$T$ linear behavior of $C_p/T = \gamma + \beta T^2$ for YCo$_{2}$Zn$_{20}$ of Figure~\ref{heat} leads to a Sommerfeld coefficient $\gamma = 18(3)$~mJ/mol~K$^2$ and a Debye temperature $\Theta_D = 370(7)$~K.
In the Fermi liquid model the Sommerfeld coefficient is given by $\gamma = (2/3)\pi^2 k_B^2 \eta_F$, where $\eta_F$ is the total density of states (DOS) per formula unit (f.u.), spin and eV at the Fermi level.
Thus, we estimate $\eta_F =3.8(8)$~states/f.u., spin and eV for YCo$_{2}$Zn$_{20}$.

A Pauli-like paramagnetic susceptibility for YCo$_{2}$Zn$_{20}$ can be estimated at high-$T$ (310~K). The data in the inset of Figure~\ref{heat} presents, after correction by the core diamagnetism of YCo$_{2}$Zn$_{20}$ ($\chi_{dia}=-2.3 \times 10^{-4}$~emu/mol), a lower limit magnetic susceptibility value at $T\approx 310~$K of $\chi_P=\chi_0-\chi_{dia}=0.312 \times 10^{-3}$~emu/mol which is slightly smaller than the value reported by Jia \emph{et al.} \cite{Jia3} Notice that the diamagnetism of the cage structure has not been considered.

Once again within the Fermi liquid model, the Pauli-like paramagnetic susceptibility is given by $\chi_P = 2\mu_B^2 \eta_F$.
Then, using the experimental Pauli paramagnetic susceptibility for YCo$_{2}$Zn$_{20}$ we estimate $\eta_F = 4.8(9)$~states/f.u., spin and eV as a lower limit for the DOS.
However, one has to consider the possibility of an exchange-enhanced magnetic susceptibility, i.e., $\chi = \chi_0/(1- \xi)$, where $\xi$ accounts for the electron-electron exchange enhancement.
Nevertheless, the estimated value of $\eta_F = 4.8(9)$~states/f.u., spin and eV is, within the accuracy or our experiments, comparable to the value obtained from the Sommerfeld coefficient.
Thus, in our analysis we shall ignore the electron-electron exchange enhancement for YCo$_{2}$Zn$_{20}$.

\begin{table}
\centering
\caption{Gd concentrations, $g$-values, residual linewidths, $a$, and thermal broadening of the linewidths, $b$, for the Y$_{1-x}$Gd$_x$Co$_2$Zn$_{20}$ system.}
\label{tab1}
\vspace{+0.5cm}
\begin{tabular}{|c||c||c||c|}
 \hline
 Conc.   & \emph{g}-value     & Measured \emph{b}   &   Calculated \emph{b}  \\
\hline
\emph{x}  &   (10 K)  &  (Oe/K)  &  (Oe/K)\\
  \hline
 0.002& 2.012(1) & 8.6(6)  &  8.4(3) \\
 \hline
 0.005& 2.006(1) & 7.1(6)   &   4.3(3)\\
 \hline
 0.085& 1.994(1) & 7.2(6)   &    0.02(3) \\
 \hline
 0.21& 1.992(1)   & 6.1(6)   &   0.02(3)\\
 \hline
 0.56& 1.991(2)   & 4.8(6)   &   0.08(5)\\
\hline
 1.00& 1.991(2)   & 1.9(6)    &   0.09(5)\\
\hline
\end{tabular}
\end{table}

From the high-$T$ data of Figure~\ref{suscept} (100~K - 300~K) for GdCo$_{2}$Zn$_{20}$ we obtain a small negative value for the Curie-Weiss temperature, $\theta_{C}$ = -0.7(4)~K, as expected for this antiferromagnetic material. 
This is in contrast to the small positive value reported in previous studies.\cite{Jia2}  

In order to complement the experimental data analysis and provide details about the electronic structure, we have performed first-principles band structure calculations for the pure compounds GdCo$_2$Zn$_{20}$ and YCo$_2$Zn$_{20}$. 
The calculated lattice constant of the fully relaxed crystal structures are 13.7738 \AA~ and 13.7564 \AA~ for GdCo$_2$Zn$_{20}$ and YCo$_2$Zn$_{20}$, respectively, obtaining an absolute relative error of 2\%, as it is expected from local and semi-local functionals. 
For GdCo$_2$Zn$_{20}$ we have obtained that the magnetic stabilization energy $\Delta_{\mathrm{AFM}} = E_{\mathrm{AFM}} - E_{\mathrm{FM}} = -2$ meV/unit cell, therefore the ground state is antiferromagnetic, with a calculated local magnetization of $7.03 \mu_{\mathrm{B}}$/Gd ion and zero local magnetization for the Co ions. 
On the other hand, YCo$_2$Zn$_{20}$ converged to a non magnetic ground state. 
A previous band structure calculation,\cite{Jia2} with local and semi-local functionals without including explicitly the Gd $f$-states into the valence window and with a tiny $\mathrm{k}$-point grid, found for the unrelaxed crystal structure of GdCo$_2$Zn$_{20}$ a slightly larger local magnetic moment of $7.25 \mu_{\mathrm{B}}$/Gd ion. 
However, it is worth noting that the calculation of the magnetic moment is very sensitive to the number of $\mathrm{k}$-points and those values were not fully converged.\cite{Jia2}

\begin{figure}[!ht]
\begin{center}
\includegraphics[width=85mm,keepaspectratio]{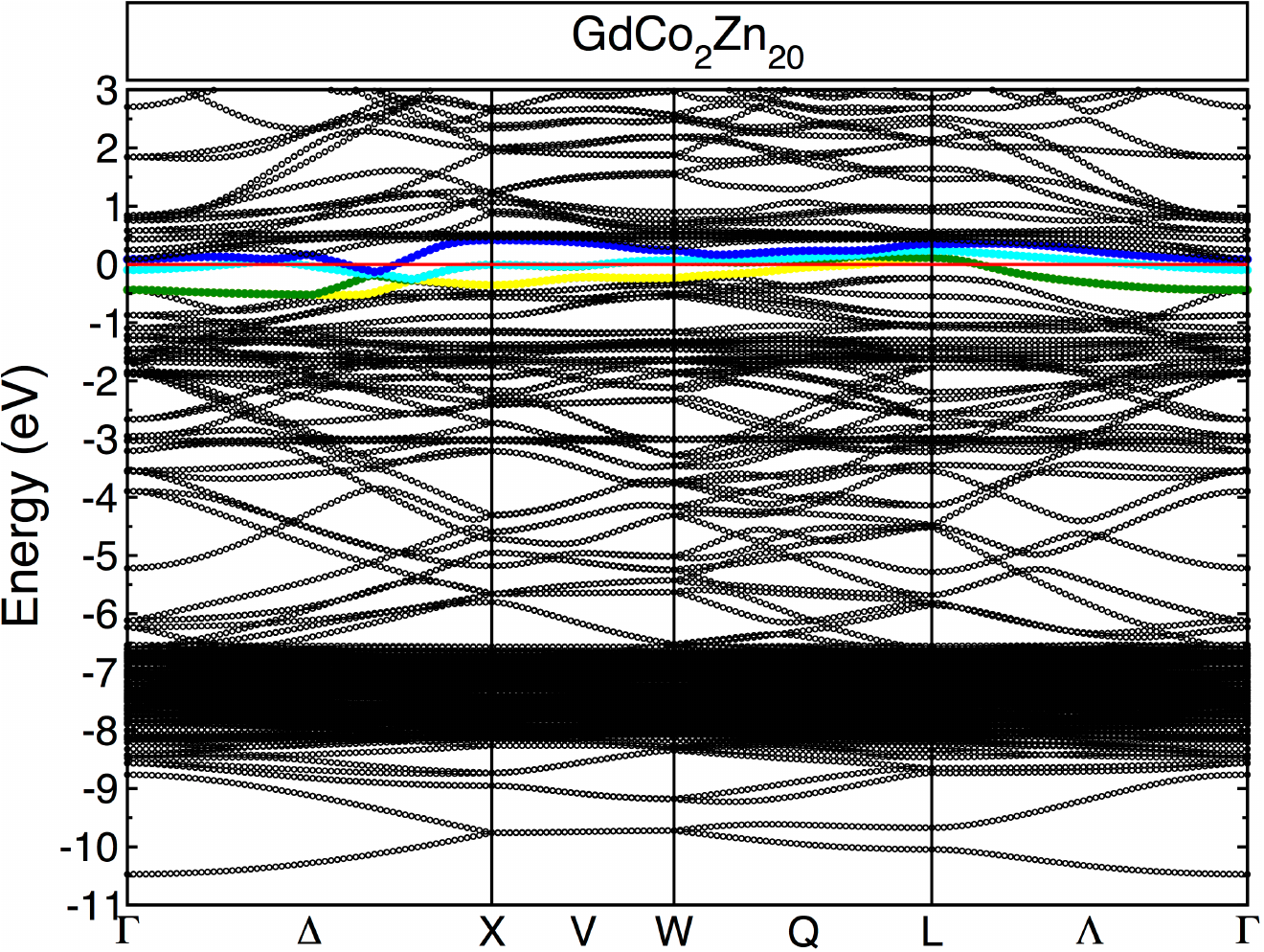}
\includegraphics[width=85mm,keepaspectratio]{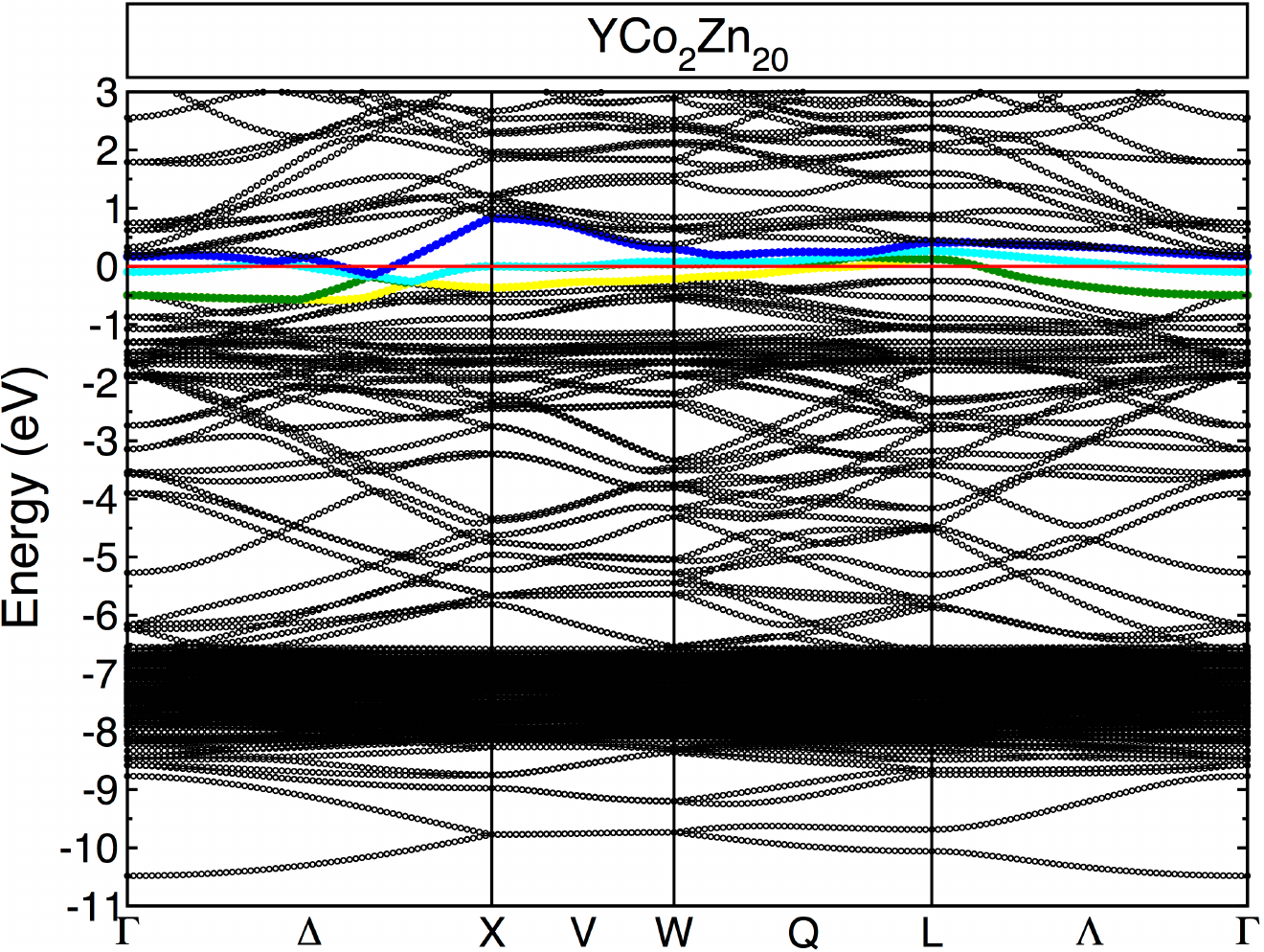}
\end{center}
\caption{Calculated dispersion relations for the GdCo$_{2}$Zn$_{20}$ (upper) and YCo$_{2}$Zn$_{20}$ (lower) systems. 
Highlighted in colors are the four conduction bands crossing the Fermi level; band 1 (yellow), band 2 (green), band 3 (cyan) and band 4 (blue).
The eigenvalues are shifted with respect to the Femi level, which is indicated by a red line.}\label{bandYGd}
\end{figure}

\begin{figure}[!ht]
\begin{center}
\includegraphics[width=85mm,keepaspectratio]{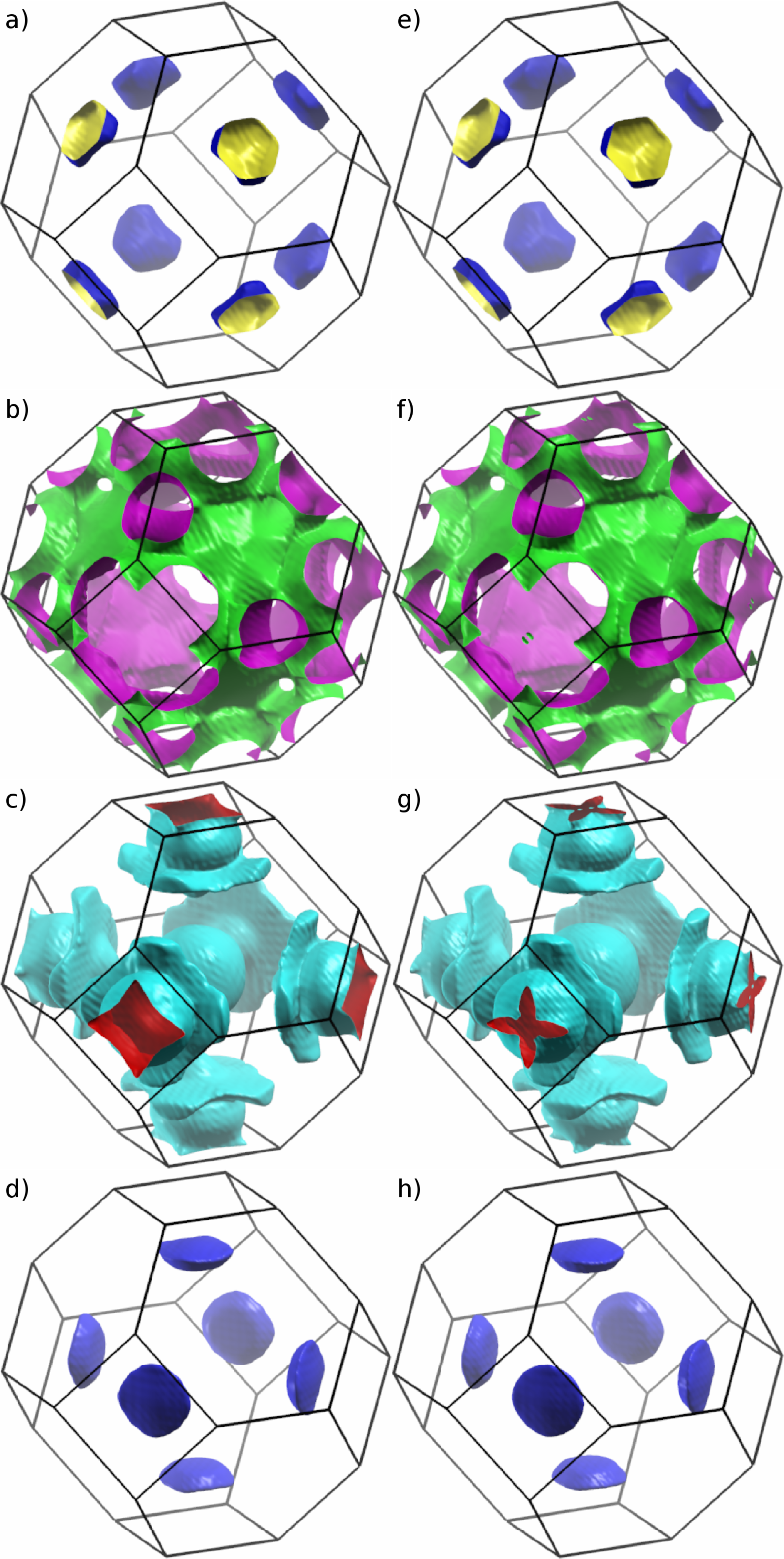}
\end{center}
\caption{Calculated Fermi surface for the GdCo$_{2}$Zn$_{20}$ and YCo$_{2}$Zn$_{20}$ systems. The four branches for GdCo$_{2}$Zn$_{20}$
corresponding to band 1, band 2, band 3 and band 4 are shown in (a), (b), (c) and (d), respectively. In the same manner, the four branches of
YCo$_{2}$Zn$_{20}$ are shown in (e), (f), (g) and (h).}\label{fermiGd}
\end{figure}

Figure~\ref{bandYGd} shows the calculated dispersion relations for these two systems. 
For GdCo$_{2}$Zn$_{20}$ the valence bands are built up mainly from Zn $d$-states and a small contribution of Zn $s$- and $p$-states between -11 and -6.5~eV for both systems.
The remaining valence bands result from the hybridization of Gd $d$- and $f$-states (the latter localized between -3.3 and -2.8~eV), Co $d$-states and Zn $s$-, $p$- and $d$-states. The conduction bands are also built up from the hybridization of Gd $d$- and $f$-states, Co $d$-states and Zn $s$-, $p$- and $d$-states, with the Gd $f$-states localized between 0.3 and 1.0~eV. 
On the other hand, the bands of YCo$_{2}$Zn$_{20}$ are built up similar to the previous system with the obvious absence of $f$-states, in this case Y $d$-states contribute mostly to the upper valence bands and the conduction bands. 
As can be observed in Fig.~\ref{bandYGd}, the topology of the bands is almost identical for the two systems, especially the four conduction bands that cross the Fermi level. 
The similarity of these four bands is more lively appreciable in the branches of the Fermi surface, as seen in Fig.~\ref{fermiGd}. 
The first branch is formed by eight pockets along the $Q$ direction of the first Brillouin zone. 
These pockets are formed from the contributions of Gd(Y) $d$-states and Co $d$-states. 
The second branch of the Fermi surface has eight connected structures along the $\Delta, V, Q$ and $\Lambda$ directions that resemble a six-arm starfish. 
These starfish are built up from Gd(Y) $d$-states, Co $d$-states and Zn $p$ and $d$-states. 
The third branch has one sphere at the center of the first Brillouin zone, which is made up from Zn $s$-states, and six structures that have the appearance of mushrooms with the stem along the $\Delta$ direction. 
These six mushrooms are formed from Gd(Y) $d$-states and Co $d$-states. 
The fourth branch is constituted of six lenses with their surfaces perpendicular to the $\Delta$ direction. 
These lenses are built up from Co $d$-states and Zn $p$-states.

\begin{figure}[!ht]
\begin{center}
\includegraphics[width=85mm,keepaspectratio]{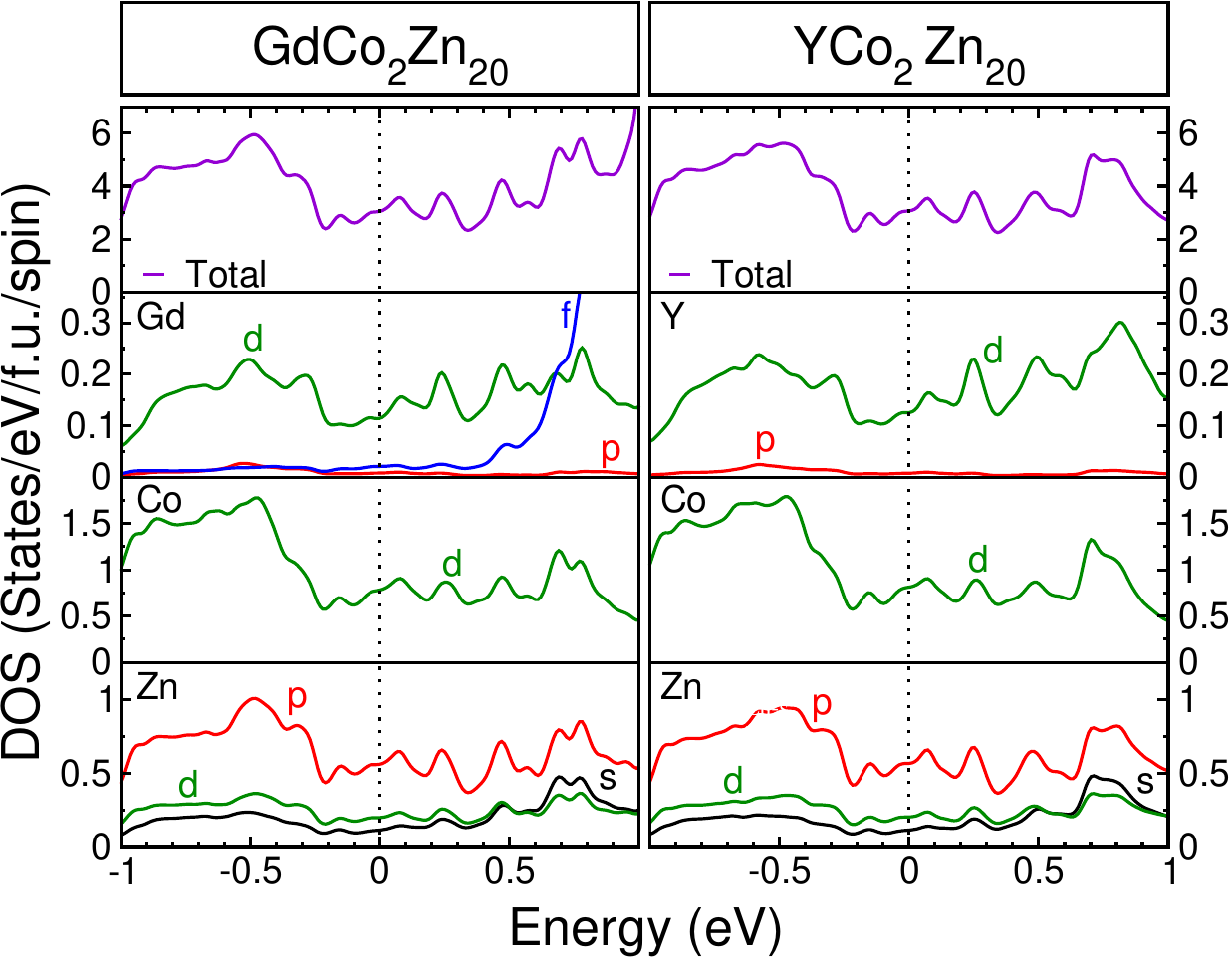}
\end{center}
\caption{ Calculated total and partial DOS for the GdCo$_{2}$Zn$_{20}$ and YCo$_{2}$Zn$_{20}$ systems. The Fermi level is indicated by
a dotted line.}\label{DOS}
\end{figure}

For GdCo$_2$Zn$_{20}$ our calculations estimate a total DOS at the Fermi level of 3.03(1)~states/f.u., spin and eV (Fig.~\ref{DOS}).
Similarly, a total DOS at the Fermi level of 3.02(1)~states/f.u., spin and eV was estimated for YCo$_2$Zn$_{20}$ (Fig.~\ref{DOS}), which is comparable to the above values obtained experimentally. 
Also, our calculated values are in relatively good agreement with a previous first-principles study with a much less dense $\mathrm{k}$-point grid.\cite{Jia2}
These results support the previous statement that YCo$_2$Zn$_{20}$ may be considered as an intermetallic compound with negligible electron-electron correlations, i.e, $\chi = \chi_0/(1- \xi)$ with $\xi << 1$.

With all these details in mind, we can now return to the ESR analysis.
The exchange interaction, $\mathcal{H}= -J_{fs}\vec{S}_f\centerdot\vec{s}_{ce}$ between the localized $4f$-electron spin of Gd$^{3+}$, $\vec{S}_f$, and the $ce$ of the YCo$_{2}$Zn$_{20}$, $\vec{s}_{ce}$, yields an ESR $g$-shift, $\Delta g$,\cite{Yosida} and thermal broadening of the linewidth, $b$, (Korringa rate)\cite{Korringa} given by:

\begin{eqnarray}
  \Delta g &=&  J_{fs}(0)\eta_{F}\nonumber\\
\end{eqnarray}

and

\begin{eqnarray}
 b = \frac{d(\Delta H)}{dT} = \frac{\pi k_B}{g\mu_B} J^2_{fs}(0) \eta^2_{F} =  \frac{\pi k_B}{g\mu_B}(\Delta g)^2,\nonumber\\
\end{eqnarray}

where $J_{fs}(0)$ is the effective exchange parameter in the absence of $ce$ momentum transfer, i.e., $\langle J_{fs}(q)\rangle_F$ = $J_{fs}(0)$,\cite{Davidov} $\eta_{F}$ is the ``bare'' density of states for one spin direction at the Fermi surface, $k_{B}$ is the Boltzmann constant, $\mu_{B}$ is the Bohr magneton, and $g$ is the Gd$^{3+}$ $g$-value. 

The above equations are normally used in the analysis of the ESR data in the limit of very diluted rare earths and concentration-independent $g$ and $b$ parameters, i.e., in the \emph{non-bottleneck} regime, absence of $ce$ momentum transfer and single band compounds.\cite{Davidov2}
Using the data in Table I and Eq. 3 one can observe that the correlation between $g$-shift, $\Delta g$, and $b$ is only verified for the lowest Gd concentration samples.
Moreover, Fig.~\ref{linewidth}, Fig.~\ref{deltag} and Table~\ref{tab1} show clear concentration-dependent ESR parameters, so our data must be analyzed in a different manner.

The change of the Gd$^{3+}$ $g$-shift from positive values (at low Gd concentrations) to negative ones (at high Gd concentrations) and the $x$-dependence of the Gd$^{3+}$ ESR thermal broadening of the linewidth, $b$, (see Fig.~\ref{dysonian} and Fig.~\ref{deltag}) lead us to conclude that the relaxation of the Gd$^{3+}$ ions to the lattice is processed via an exchange interaction, $J(\vec{S}_f\centerdot\vec{s}_{ce})$, between the Gd$^{3+}$ localized magnetic moment and different types of \emph{ce} at the Fermi level.

In a multiband approximation the $g$-shift, $\Delta g$, and thermal broadening of the linewidth, $b$, are given by:

\begin{eqnarray}
  \Delta g &=& \Delta g_{fs} + \Delta g_{fp} + \Delta g_{fd}\nonumber\\
  \\
   &=& J_{fs}(0)\eta_{F_s} - J_{fp}(0)\eta_{F_p} + J_{fd}(0)\eta_{F_d}\nonumber
\end{eqnarray}

and

\begin{eqnarray}
 b &=& \frac{\pi k_B}{g\mu_B} [F_s\Delta g^2_{fs} + F_p\Delta g^2_{fp} + F_d\Delta g^2_{fd}]\nonumber\\
 \\
 &=&\frac{\pi k_B}{g\mu_B} [F_s\langle J^2_{fs}(q)\rangle_F \eta^2_{F_s} + F_p J_{fp}^2(0) \eta^2_{F_p} + F_d J_{fd}^2(0) \eta^2_{F_d}],\nonumber
\end{eqnarray}

\vspace{2pc}

where $k_B$ is the Boltzmann constant, $\mu_B$ the Bohr magneton, and $g$ the Gd$^{3+}$ $g$-value; $J_{fi}(0)\,(i= s, p, d)$ are the effective $q = 0$ components of the exchange interaction between the Gd$^{3+}$ 4\emph{f} magnetic moment and the \emph{s}-, \emph{p}- and \emph{d}-type \emph{ce}; $\eta_{F_i}\,(i = s, p, d)$ the partial \emph{bare} DOS (states/f.u., spin and eV) at the Fermi level of the \emph{s}-, \emph{p}- and \emph{d}-type \emph{ce}; $\langle J^2_{fs}(q)\rangle_F$ is the average over the Fermi surface of the square of the $q$-dependent effective exchange parameter in the presence of \emph{ce} momentum transfer, $q = |\vec{k}_{out} - \vec{k}_{in}|$, i.e., $\langle J_{fs}(q)\rangle_F \neq J_{fs}(0)$;\cite{Davidov} $F_s = 1$, $F_p = 1/3$ and $F_d = 1/5$ are factors associated with the orbital degeneracy of the unsplit (no crystal field effects) bands at the Fermi level, respectively.
The $q$-dependence of the exchange interaction with the \emph{p}-type and \emph{d}-type \emph{ce} will be considered constant over the Fermi surface, i.e., $\langle J_{fp,d}(q)\rangle_F = J_{fp,d}(0)$ (see below).

In Eq. 4 we have considered that the contribution to $\Delta g$ due to the exchange interaction with \emph{s}- and \emph{d}-type \emph{ce} are positive (atomic-like) and that with \emph{p}-type \emph{ce} is negative (covalent-like).\cite{Davidov}

Due to the strong spin-orbit coupling of \emph{p}- and \emph{d}-type \emph{ce} compared to that of the \emph{s}-type \emph{ce}, we assume that only the \emph{s}-type \emph{ce} are capable of experiencing the \emph{bottleneck} effect.
Hence, we can consider that the contribution to the ESR parameters, $g$-shift and $b$, of the \emph{s}-type \emph{ce} are negligible in the highly concentrated samples.
With these assumptions, for $x = 1$ (GdCo$_{2}$Zn$_{20}$, extreme \emph{bottleneck}) Eq. 4 reduces to

\begin{eqnarray}
 \Delta g = -0.002(2) = -J_{fp}(0)\eta_{F_p} + J_{fd}(0)\eta_{F_d},
\end{eqnarray}

and Eq. 5 reduces to (see Fig.~\ref{DOS}b):

\begin{eqnarray}
 b = 1.9(6)\,\textrm{Oe/K} = \frac{\pi k_B}{g\mu_B} [F_p J_{fp}^2(0)\eta^2_{F_p} + F_d J_{fd}^2(0)\eta^2_{F_d}].
\end{eqnarray}

\vspace{2pc}

In general, the $q$-dependence of the exchange parameters $J_{f,i}$ $(i = s,p,d)$ cannot be disregarded, but in the case in which the relaxation rate, $b$, scales or is slightly larger than the expected value from the $g$-shift ($b=(\pi k_B /g\mu_B) (\Delta g)^2$) we can neglect the $q$-dependence of the $J_{fp}$ and $J_{fd}$ exchange parameters.
In our case, we have that 1.9~Oe/K~$\geq (\sim2.34 \times10^4$~Oe/K)$\times(-0.002)^2 \approx 0.1$~Oe/K in agreement with Table I.

From Fig.~\ref{DOS} for GdCo$_{2}$Zn$_{20}$, we have $\eta_{F_d}$ = 1.09(1) states/f.u., spin and eV and $\eta_{F_p}$ = 0.59(1) states/f.u., spin and eV.
Then, using Eqs.~6 and 7 we estimate $J_{fd}(0) = 10(5)$~meV and $J_{fp}(0) = 22(6)$~meV.

Conversely, in the \emph{not bottlenecked} regime (lowest Gd concentration, $x\approx 0.002$), and from Fig.~\ref{DOS} for YCo$_{2}$Zn$_{20}$, Eqs. 4 and 5 reduce to:

\begin{eqnarray}
 \Delta g = 0.019(2) = J_{fs}(0)\eta_{F_s} - 0.002(2),
\end{eqnarray}

\begin{eqnarray}
 b = 8.6(6)\,\textrm{Oe/K} = \frac{\pi k_B}{g\mu_B} [F_s\langle J^2_{fs}(q)\rangle_F\eta^2_{F_s}] + 1.9(6).
\end{eqnarray}

\vspace{2pc}

Again from Eqs. 8 and 9 and using $\eta_{F_s}$ = 0.13(1) states/f.u., spin and eV, we obtain $J_{fs}(0) = 167(7)$~meV and $\langle J^2_{fs}(q)\rangle_F^{1/2} = 18(5)$~meV.
Comparing this value with those for $J_{fp}(0)$ and $J_{fd}(0)$, we find that in these compounds the polarization component of the exchange parameter $J_{fs}(0)$ is an order of magnitude larger. 

Thus, in order to describe the antiferromagnetic ordering of GdCo$_{2}$Zn$_{20}$, it should be more appropriate to use an RKKY approach that considers only the exchange parameter $J_{fs}(0)$ (due to the delocalized nature of the $s$-type electrons compared with $p$ and $d$-type) rather than the Campbell model,\cite{Campbell,Shuo} that considers the $J_{fd}(0)$ term as the most important one. 
Therefore our obtained value in ESR analysis for $J_{fs}(0)$ can be used to establish a correlation with the RKKY interaction, which is only valid in very dilute magnetic systems as this family of compounds.

In general, the RKKY interaction depends strongly on the Fermi surface and can have different analytical forms for each case.
Finding a suitable expression for a real material with a complex Fermi surface is thus expected to be an almost impossible task. 
The equation that represents a generalized form of the RKKY interaction is given by\cite{Jensen}:

\begin{eqnarray}
 J_{RKKY} \sim J_{fs}^{2}\sum\limits_{\bold{k,q}}\frac{f_{\bold{k}}-f_{\bold{k+q}}}{\epsilon (\bold{k+q})-\epsilon (\bold{k})},
\end{eqnarray}

where $f_{\bold{k}}=\Theta(k_{F}-|\bold{k}|)$ and $f_{\bold{k+q}}=\Theta(k_{F}-|\bold{k+q}|)$ are the step functions that come from the non-zero matrix elements, according to the second-order perturbation theory treatment of the second quantized Heisenberg hamiltonian, between a localized and an itinerant electron spin coupled by $J_{fs}$. 
The term $\epsilon (\bold{k+q})-\epsilon (\bold{k})$ corresponds to the energy diference between the ground state and the excited state. 
This function is known as a Lindhard function that appears in the generalized form of the magnetic susceptibility\cite{Jensen}. 
Transforming the two summations into integrals, one may reach the actual expression for the RKKY interaction. 
Our calculated Fermi surface, shown in Fig.~\ref{fermiGd}, evidences a rather complex dispersion relation for the GdCo$_{2}$Zn$_{20}$ compound, so obtaining an exact analytical solution from Eq. 10 is not possible.

However, within a simple Fermi gas model ($\epsilon(\bold{k}) \sim k^{2}$) the problem is simplified and more accessible. 
In this approximation the analytical expression of the RKKY interaction for the effective coupling between two lattice localized spins is given, in terms of $k_{F}$, by\cite{Coey}:

\begin{eqnarray}
  J_{eff} \approx  \frac{9\pi (J_{fs}(0))^{2} \nu^{2} F(r)}{64 E_{F}(2 k_{F})^{4}},\nonumber\\
\end{eqnarray}

where $\nu$ is the number of conduction electrons per atom, $E_{F}$ the Fermi energy, $J_{fs}(0)$ the coupling constant between the Gd$^{3+}$ spin $S_f$ and the conduction electron spin $s_{ce}$, ($\mathcal{H}= -J_{fs}(0)\vec{S_f}\centerdot\vec{s_{ce}}$), and $F(r)$ is the RKKY function given by:

\begin{eqnarray}
 F(r) = \frac{\sin(2 k_{F} r)-2 k_{F} r \cos(2 k_{F} r)}{r^{4}}.
\end{eqnarray}

Hence, in the GdCo$_{2}$Zn$_{20}$ compound, it is possible to estimate the effective exchange parameter, $J_{eff}$ (Eq. 11), due to the coupling between two nearest Gd$^{3+}$ neighbors via the damped spin polarization of only the $s$-type \textit{ce}, as it can observed in the third branch of the
Fermi surface (Fig.~\ref{fermiGd} (c,g)).

\begin{figure}[!ht]
\begin{center}
\includegraphics[width=85mm,keepaspectratio]{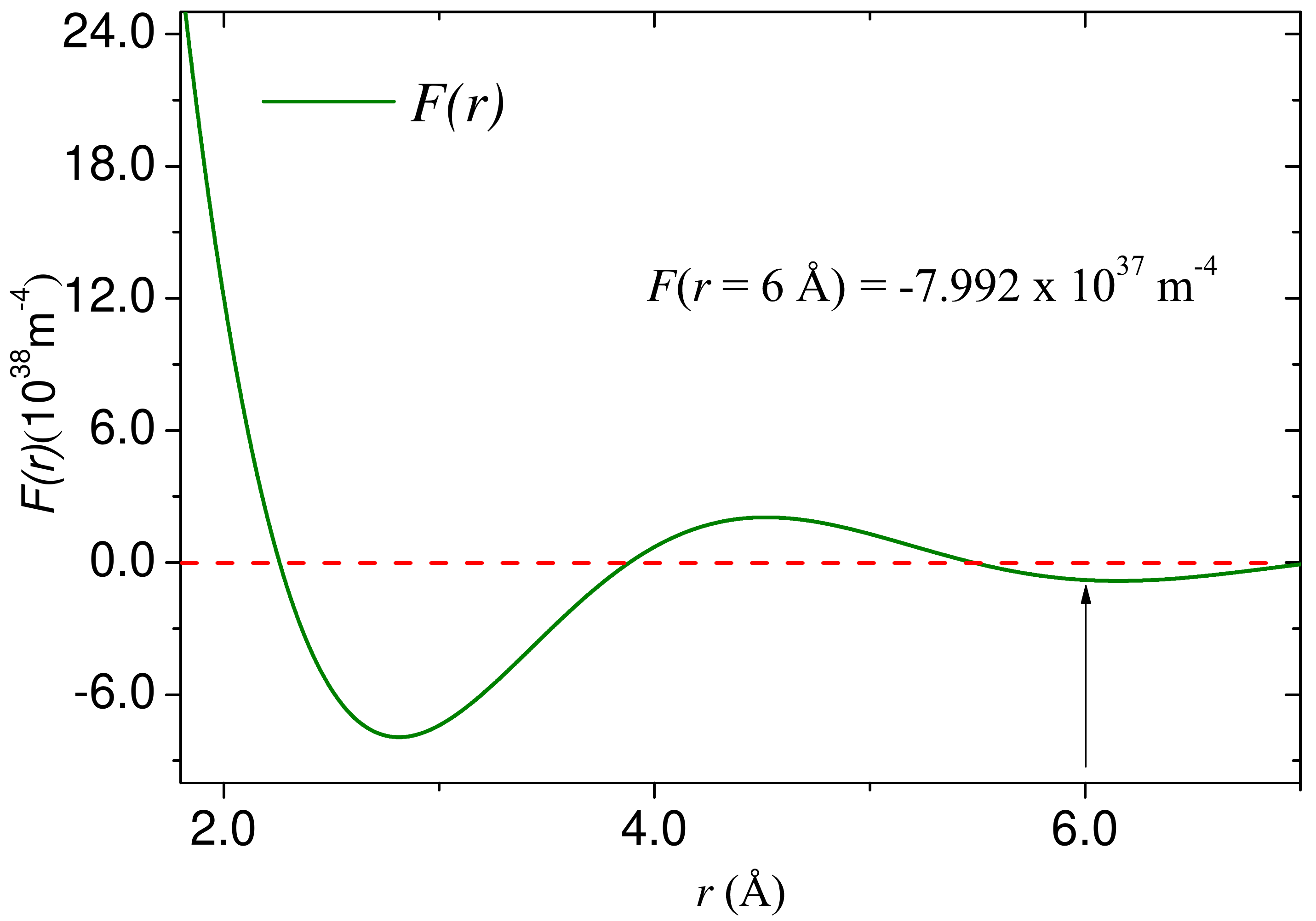}
\end{center}
\caption{RKKY oscillatory coupling function $F(r)$ as a function of distance $r$ between Gd ions.
The arrow indicates the Gd-Gd separation in the GdCo$_{2}$Zn$_{20}$ system.}\label{RKKY}
\end{figure}

For GdCo$_{2}$Zn$_{20}$ the following band structure parameters were determined from the DFT calculations: $E_{F}$ = 3.77 eV with $k_{F} = 0.99478 \times 10^{10}$~m$^{-1}$ ($k_{F}$ = $\frac{1}{\hslash}$$\sqrt{2m_{e}E_{F}}$) and $\nu = 1.23$~c.e/atom.
Notice that the value of $k_{F} = 0.99478 \times 10^{10}$~m$^{-1}$ is comparable to those reported for uncorrelated simple metals such as Cu, Ag and Au.\cite{Aschcroft}
Then, the amplitude of the RKKY function $F(r)$ at $r = 6.0$~\AA ~(nearest Gd-Gd neighbors) gives $F$(6.0~\AA) $=-7.99\times 10^{37}$~m$^{-4}$ (Fig.~\ref{RKKY}) and, consequently, $J_{eff} = -25.2(7)\times 10^{-4}$~meV, a negative value as expected for antiferromagnetic ordering.

\
Now, a \emph{microscopic} Curie-Weiss temperature estimation can be made from this result:\cite{Coey}

\begin{eqnarray}
 \theta_{C} = \frac{2 Z J_{eff}S(S+1)}{3 k_{B}},
\end{eqnarray}
\
with $Z = 4$ (Gd nearest neighbors in GdCo$_{2}$Zn$_{20}$) and $S=7/2$ for Gd$^{3+}$.
We obtain $\theta_{C} = -1.2(2)$~K which is, within the accuracy of our experiments, in very good agreement with the \emph{bulk} estimation (Fig.~\ref{suscept}) extracted from the magnetic susceptibility measurements.

\section{Conclusions}

Our experimental ESR results of Y$_{1-x}$Gd$_{x}$Co$_{2}$Zn$_{20}$ (0.002 $\lesssim x \leq $ 1.00) were analyzed within a multiband model of \emph{ce} ($s$-, $p$- and $d$-type) where, via the Gd concentration, the system was tuned from an \textit{non-bottleneck} regime (Y$_{0.998}$Gd$_{0.002}$Co$_{2}$Zn$_{20}$) to a \emph{bottleneck} regime (GdCo$_{2}$Zn$_{20}$).
The combination of ESR results with those of heat capacity, magnetic susceptibility and band structure calculations allowed us to estimate the polarization component of the exchange parameters, $J_{fi}(0)\,(i = s, p, d)$.
Besides, by the assumption that only the \emph{s}-type of \emph{ce} can experience the \emph{bottleneck} effect, due to their relatively weak spin-orbit coupling, we found that the average over the Fermi surface of the exchange parameter, associated to the \emph{ce} momentum transfer, is different from the exchange parameter leading to local polarization effects, i.e., $\langle J_{fs}(q)\rangle_F \neq J_{fs}(0)$.

The exchange parameters obtained with this multiband scenario revealed that $J_{fs}(0)$ is dominant over $J_{fp}(0)$ and $J_{fd}(0)$. 
This allowed a tractable RKKY description for the antiferromagnetic behavior of the GdCo$_{2}$Zn$_{20}$ compound.
Despite the fact that these compounds are structurally complex, we found that under certain reasonable approximations and using the combination of different experimental results with DFT calculations, the RKKY approach gave a very good quantitative description of the magnetic interaction, as expected for a naturally diluted structure of rare-earth ions. 
This was confirmed by the reasonably accurate prediction of the Curie-Weiss temperature in terms of microscopic parameters.

With the resulting establishment of GdCo$_{2}$Zn$_{20}$ as a model RKKY system, we expect that this work can provide key reference elements to help understand the behaviors of related materials such as RFe$_{2}$Zn$_{20}$ and YbT$_{2}$Zn$_{20}$, with their more complex and remarkable electronic and magnetic properties.

\vspace{2pc}

\begin{acknowledgments}
This work was supported by Brazilian agencies FAPESP (Grant Nos. 2011/19924-2, 2012/17562-9), CNPq, FINEP and CAPES. We thank R. A. Ribeiro for sample preparation support and P. G. Pagliuso for fruitful discussions.
JMOG would like to thank CODI-Vicerrector\'ia de Investigaci\'on-Universidad de Antioquia (Estrategia de Sostenibilidad 2014--2015).
\end{acknowledgments}

\FloatBarrier

\end{document}